\documentclass[lettersize,journal]{IEEEtran}
\usepackage{amsmath,amsfonts}
\usepackage{algorithmic}
\usepackage{algorithm}
\usepackage{array}
\usepackage[caption=false,font=normalsize,labelfont=sf,textfont=sf]{subfig}
\usepackage{textcomp}
\usepackage{stfloats}
\usepackage{url}
\usepackage{verbatim}
\usepackage{graphicx}
\usepackage{cite}
\usepackage{nomencl}
\usepackage{multirow}
\usepackage{amsthm}
\usepackage{booktabs}
\usepackage{todonotes}

\newtheorem{lemma}{Lemma}

\usepackage{amssymb} 
\usepackage{xcolor} 

\makenomenclature
\usepackage{etoolbox}
\renewcommand\nomgroup[1]{%
  \item[\bfseries
  \ifstrequal{#1}{A}{Indices}{%
  \ifstrequal{#1}{B}{Sets}{%
  \ifstrequal{#1}{C}{Constants}{%
  \ifstrequal{#1}{D}{Variables}{}}}}%
]}
\begin{document}

\title{Optimal Trading of a Charging-Station Company in Auction Markets for Electricity}

\author{Farnaz Sohrabi, Mohammad Rohaninejad, Mohammad Reza Hesamzadeh,~\IEEEmembership{Senior Member,~IEEE},~J\'{u}lius~Bem\v{s}}

\maketitle

\begin{abstract}
This paper addresses a charging-station company (Chargco) for electric and hydrogen vehicles. The optimal trading of the Chargco in day-ahead and intraday auction markets for electricity is modeled as a stochastic Mixed-Integer Quadratic Program (MIQP). We propose a series of linearization and reformulation techniques to reformulate the stochastic MIQP as a mixed-integer linear program (MILP). To model stochasticity, we utilize generative adversarial networks to cluster electricity market price scenarios. Additionally, a combination of random forests and linear regression is employed to model the relationship between Chargco electricity and hydrogen loads and their selling prices. Finally, we propose an Improved L-Shaped Decomposition (ILSD) algorithm to solve our stochastic MILP. 
Our ILSD algorithm not only addresses infeasibilities through an innovative approach but also incorporates warm starts, valid inequalities and multiple generation cuts, thereby reducing computational complexity. Numerical experiments illustrate the Chargco trading using our proposed stochastic MILP and its solution algorithm. 
\end{abstract}


\printnomenclature
\nomenclature[A]{$t', t$}{Index for day-ahead and intraday intervals.}
\nomenclature[A]{$\omega_{1}, \omega_{2}$}{Index for day-ahead and intraday scenarios.}
\nomenclature[A]{$j$}{Index for bidding price steps.}
\nomenclature[A]{$\omega$}{Index for scenarios.}

\nomenclature[C]{$\pi_{\omega}$}{Probabilities associated with scenarios.}
\nomenclature[C]{$\xi_{t'\omega_{1}}^{d}, \xi_{t \omega_{2}}^{i}$}{Day-ahead and intraday electricity prices.}
\nomenclature[C]{$\eta^{b}, \eta^{h}, \eta^{e}$}{Efficiency of battery, tank, and electrolyzer.}
\nomenclature[C]{$\underline{b}^{l}, \overline{b}^{l}$}{Min and max for battery level capacity.}
\nomenclature[C]{$\underline{b}^{c}, \overline{b}^{c}$}{Min and max for battery charging capacity.\footnote{Min: minimum, Max.: Maximum}}
\nomenclature[C]{$\underline{b}^{d}, \overline{b}^{d}$}{Min and max for battery discharging capacity.}
\nomenclature[C]{$\underline{e}^{p}, \overline{e}^{p}$}{Min and max for electrolyzer capacity.}
\nomenclature[C]{$y_{j}$}{Electricity bidding price steps.}
\nomenclature[C]{$\underline{h}^{l}, \overline{h}^{l}$}{Min and max for tank capacity.}
\nomenclature[C]{$\underline{h}^{c}, \overline{h}^{c}$}{Min and max for tank charging capacity.}
\nomenclature[C]{$\underline{h}^{d}, \overline{h}^{d}$}{Min and max for tank discharging capacity.}
\nomenclature[C]{$\mathrm{H}$}{Heating value of hydrogen.}

\nomenclature[D]{$l_{t}^{e}, l_{t}^{h}$}{Electricity and hydrogen loads.}
\nomenclature[D]{$p_{t}^{e}, p_{t}^{h}$}{Electricity and hydrogen selling prices.}
\nomenclature[D]{$m_{t\omega_{1}}^{d}, m_{t\omega}^{i}$}{Day-ahead and intraday bought power.}
\nomenclature[D]{$v_{t\omega}^{e}, v_{t\omega}^{h}$}{Dedicated electricity and hydrogen for loads.}
\nomenclature[D]{$b^{c}_{t\omega}, b^{d}_{t\omega}$}{Charged and discharged power from battery.}
\nomenclature[D]{$e_{t\omega}^{p}$}{Consumed power by electrolyzer.}
\nomenclature[D]{$b^{l}_{t\omega}, h^{l}_{t\omega}$}{Power and hydrogen level in battery and tank.}
\nomenclature[D]{$x_{j t'}^{d}, x_{j t \omega_{1}}^{i}$}{Bid volume in day-ahead and intraday markets.}
\nomenclature[D]{$h^{c}_{t\omega}, h^{d}_{t\omega}$}{Charged and discharged hydrogen from tank.}
\nomenclature[D]{$u^{b}_{t\omega}, u^{h}_{t\omega}$}{Operation mode for battery and tank.}

\section{Introduction}
\IEEEPARstart{T}{he} use of fossil fuels has a negative effect on the environment and depletes these resources \cite{dahiwale2024comprehensive}.
Transportation accounts for roughly 20\% of global $\text{CO}_{2}$ emissions, with three-quarters of this attributed to road transport \cite{owid-co2-emissions-from-transport}.
This has led to a growing interest in battery electric and hydrogen fuel cell vehicles as environmentally friendly alternatives \cite{sohrabi2023coordinated}.
As a result, innovative solutions such as combined electric and hydrogen charging stations are emerging to meet the growing demand for sustainable transportation. However, research on these systems is still limited, creating a gap in our understanding, particularly regarding their participation in electricity markets. It is essential to explore how these stations can effectively engage in electricity markets, develop effective solution methods, and manage the uncertainties that arise in this context. This paper addresses these challenges by investigating strategies for integrating these stations into the electricity market. We focus on key aspects, including mathematical modeling, linearization techniques, and the implementation of adaptable solution algorithms capable of responding to the dynamic nature of energy trading and its associated uncertainties.


\subsection{Literature review}
Authors in \cite{shoja2022sustainable} propose a day-ahead optimization framework for sustainable energy supply to an electric and hydrogen vehicle refueling station, utilizing a hybrid multi-objective strategy to handle uncertainties. Similarly, Panah et al. \cite{panah2021charging} tackle the challenge of developing a charging station that meets both hydrogen and electricity needs, serving public transportation systems alongside private vehicles. In off-grid areas, Xu et al. \cite{xu2020optimal} exclusively rely on solar photovoltaics for electric vehicle charging and hydrogen refueling.
An on-grid station is proposed in \cite{xu2022robust} to support hydrogen and electric vehicles, incorporating a hybrid stochastic and distributionally robust optimization method to address uncertainties in energy management.
An autonomous hybrid charging station powered by a photovoltaic system for charging hydrogen and electric vehicles, which addresses inherent demand uncertainties using information-gap decision theory (IGDT), is presented in \cite{sriyakul2021risk}.
A real-time energy management strategy for intelligent solar parking lots that charge both electric and hydrogen vehicles is presented in \cite{guo2022energy}, addressing challenges posed by uncertainties through a deep reinforcement learning model.
Elmasry et al. present an integrated electricity-hydrogen system with multi-level hydrogen storage and solar PV that supplies electric and fuel cell vehicles at three charging points: an electric vehicle station, a fuel cell car refueling station, and a fuel cell truck refueling station, with the cost function representing daily energy costs \cite{elmasry2024electricity}.
In \cite{cai2023hierarchical}, a bi-level coordination optimization model is introduced for an integrated energy system and a hybrid charging station, with the aim of maximizing social benefits while meeting the energy requirements of environmentally friendly vehicles for both electricity and hydrogen.
The energy needs of an off-grid refueling station for electric and hydrogen vehicles, along with a temporary residence, are designed and analyzed in \cite{dastjerdi2023transient} while renewable energy sources such as wind and solar power are also considered.
A low-carbon energy management model for a microgrid that integrates hydrogen, heat, and power systems, incorporating charging stations to simultaneously meet the energy needs of both hydrogen and electric vehicles, is presented in \cite{mansour2021multi}.
An economically optimal energy management model for a charging station that utilizes a photovoltaic system and an electrolyzer to charge both electric and hydrogen vehicles is proposed in \cite{cciccek2022optimal}. This study employs a mixed-integer linear programming approach.
In \cite{schroder2020optimization}, a grid-connected photovoltaic system for charging electric vehicles and refueling fuel cell vehicles was examined through numerical simulations. The proposed system is optimized using genetic algorithm to identify cost-effective component sizes and energy management strategies.

The existing literature has not thoroughly examined the trading activities of charging station owners in electricity markets, presenting a research gap. While various studies have explored the optimization and operational strategies of charging stations, the specific dynamics of how charging station owners engage in electricity trading remain largely unexplored. This oversight limits our understanding of the potential economic benefits and challenges that charging station operators may face in these markets. As the electrification of transport systems continues to evolve, it is crucial to investigate how these stakeholders can effectively participate in electricity trading to enhance their profitability and sustainability. Addressing this gap will provide valuable insights into the strategic decisions of charging station owners, potentially leading to more robust frameworks that support the energy transition.

\begin{table}[!ht]
\centering
\caption{Comparison of Selected Papers on Energy Management for Hydrogen-Electric Charging Stations}
{
\begin{tabular}{|>{\centering\arraybackslash}p{1.2cm}|>{\centering\arraybackslash}p{0.4cm}|c|c|>{\centering\arraybackslash}p{0.5cm}|>{\centering\arraybackslash}p{0.5cm}|c|}
\hline
\textbf{Ref.} & \textbf{Lin.} & \textbf{Uncert.} & \textbf{Decomp.} & \textbf{DAM} & \textbf{IDM} & \textbf{Bid.} \\ 
\hline
\cite{shoja2022sustainable} & \checkmark & IGDT/Robust & $\times$ & \checkmark & $\times$ & $\times$ \\ 
\hline
\cite{panah2021charging} & $\times$ & Stoch. & $\times$ & \checkmark & $\times$ & $\times$ \\ 
\hline
\cite{xu2020optimal} & \checkmark & Stoch. & $\times$ & $\times$ & $\times$ & $\times$ \\  
\hline
\cite{xu2022robust} & $\times$ & Stoch./DRO & $\times$ &  \checkmark & $\times$ & $\times$ \\ 
\hline
\cite{sriyakul2021risk} & \checkmark & IGDT & $\times$ & $\times$ & $\times$ & $\times$ \\ 
\hline
\cite{guo2022energy} & $\times$ & Deep RL & $\times$ & $\times$ & $\times$ & $\times$ \\ 
\hline
\cite{elmasry2024electricity} & $\times$ & Stoch. & $\times$ & $\times$ & $\times$ & $\times$ \\ 
\hline
\cite{cai2023hierarchical} & $\times$ & DRO & $\times$ &  \checkmark & $\times$ & $\times$ \\ 
\hline
\cite{dastjerdi2023transient} & $\times$ & $\times$ & $\times$ & $\times$ & $\times$ & $\times$ \\ 
\hline
\cite{mansour2021multi} & \checkmark & IGDT & $\times$ & \checkmark & $\times$ & $\times$ \\ 
\hline
\cite{cciccek2022optimal} & \checkmark & $\times$ & $\times$ & $\times$ & $\times$ & $\times$ \\ 
\hline
\cite{schroder2020optimization} & $\times$ & Stoch. & $\times$ & $\times$ & $\times$ & $\times$ \\ 
\hline
This study & \checkmark & Stoch. & \checkmark & \checkmark & \checkmark & \checkmark \\ 
\hline
\end{tabular}
}
\end{table}

As innovative charging station solutions gain attention, understanding their integration into the electricity markets becomes crucial. The European energy exchange introduced diverse trading opportunities, spanning futures, spot, and balancing markets. Our focus is on the spot market, renowned for its prominence in the energy landscape \cite{narajewski2022optimal}.
In most European countries, the spot market encompasses both day-ahead and intraday markets \cite{ocker2020way}.  We focus on discrete-auction mechanisms, which entails buying and selling power one day or several hours in advance to meet the real-time demand. 

Charging station owner purchases electrical energy from the grid, but the owner's profitability is affected by the uncertainty in electricity market prices \cite{bagherzadeh2020long}.
Generative Adversarial Networks (GANs) can be utilized for clustering electricity market price scenarios from historical data without the need for explicit model specifications. It utilizes unsupervised learning to eliminate the need for time-consuming manual labeling, which is particularly impractical for large datasets \cite{chen2018model}.
The GANs, developed by Goodfellow et al. \cite{goodfellow2014generative}, consist of two neural networks engaged in a competitive game. The generator produces synthetic samples from a training dataset, while the discriminator distinguishes between fake and real samples. 

The random forest method is categorized as an ensemble method, combining multiple classifiers into a single system. It achieves higher classification accuracy than any individual classifier that could attain alone \cite{deng2017learning}. 
The main advantage of the random forest method is its independence from prior knowledge regarding the data distribution \cite{shevchik2016prediction}.
The random forest uses multiple decision trees to reduce errors. It involves randomly selecting training instances, choosing feature subsets to split nodes, and learning from random samples during training \cite{jahangiri2015applying}.

In discrete optimization, stochastic Mixed-Integer Linear Programs (MILPs) cause significant challenges due to increasing memory requirements and computational workload with scenario growth. 
Various decomposition techniques have been developed to address this issue by breaking down complex stochastic MILP problems into more manageable sub-problems \cite{moiseeva2017strategic, khastieva2021optimal}. L-shaped decomposition is one such technique that estimates the expected second-stage recourse function through optimality cuts in the first stage, employing dual solutions from the second-stage problems. This iterative process between a master problem and a sub-problem exchanges information to derive the optimal solution \cite{fazeli2020two}.

\vspace{-1mm}

\subsection{Contributions} 
The contributions of the current paper are as follows:

\textbf{First}, we develop a two-stage stochastic Mixed-Integer Quadratic Program (MIQP) for optimal trading of a charging-station company (Chargco) in Day-Ahead (DA) and Intraday (ID) markets. A series of linearization techniques is applied to approximate the MIQP as a MILP. This includes piecewise linearization method utilizing special ordered sets of type 2 (SOS2) variables and max-affine functions.

\textbf{Second}, we develop a surrogate model incorporating random forest and linear regression to enhance the model's ability to predict electricity and hydrogen load levels, along with their corresponding prices. 

\textbf{Third}, To address challenges related to scenario reduction, we utilize GANs on annual real electricity market prices.  The GANs is utilized to cluster scenarios, effectively representing uncertainties in both DA and ID electricity prices.

\textbf{Fourth}, the approximated MILP is solved using an Improved L-Shaped Decomposition (ILSD) algorithm. The original MILP is broken down into a Master Problem (MP) that optimizes hydrogen and electricity load levels, selling prices, and day-ahead bidding, followed by a Sub-Problem (SP) for intraday bidding and device operations. We demonstrate in \textbf{Lemma \ref{lem:sp}} that the SP can be reformulated as an equivalent LP. Additionally, our ILSD algorithm eliminates the need for feasibility cuts or slack variables, as shown in \textbf{Lemma \ref{lem:infs}}. The algorithm is further enhanced by warm-start techniques, valid inequalities, and multiple-cuts generation to reduce computational complexities. The effectiveness of the MILP model and its solution algorithm is illustrated through a case study, providing valuable insights for decision making in electricity markets.

The paper is structured as follows: Section \ref{Problem statement} formulates the stochastic MILP for Chargco optimal trading and explains the random forest method. Scenario generation using GANs is detailed in Section \ref{Generative adversarial network}. The improved L-shaped decomposition algorithm is discussed in Section \ref{Methodology}, followed by the case study in Section \ref{Case study}.Section \ref{Further discussions and future works} provides further discussions and future work. Further discussions and future works.Finally, Section \ref{Conclusion} concludes the paper.

\section{Profit-Maximizing Charging-Station Company for Electric and Hydrogen Vehicles} \label{Problem statement}
\begin{figure}[hbt!]
    \centering    \includegraphics[width=0.95\columnwidth]{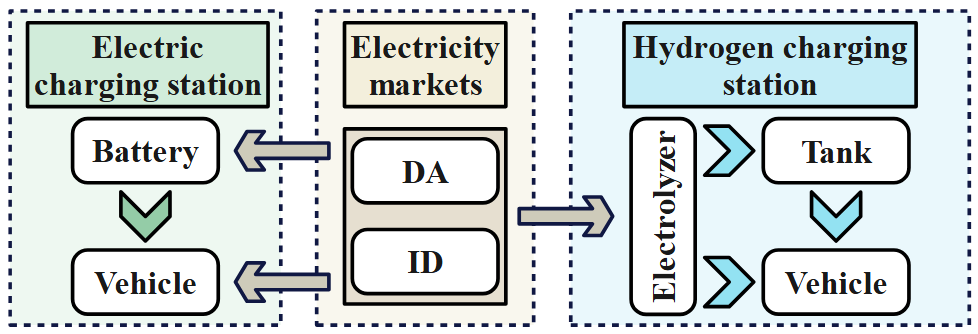}
    \caption{The Chargco owning ECSs and HCSs participating in markets }
    \label{fig:station}
\end{figure}

We consider a profit-maximizing charging-station company owning electric and hydrogen vehicle charging stations. Chargco submits bids to DA and ID markets, purchasing electricity for Electric Charging Stations (ECSs) and Hydrogen Charging Stations (HCSs), as shown in Fig. \ref{fig:station}. The purchased electricity serves three purposes: direct electric vehicle charging, charging battery storage, and hydrogen production via the electrolyzer, for storage or direct hydrogen vehicle charging.

\vspace{-1mm}

\begin{figure}[!hbt]
    \centering    \includegraphics[width=0.99\columnwidth]{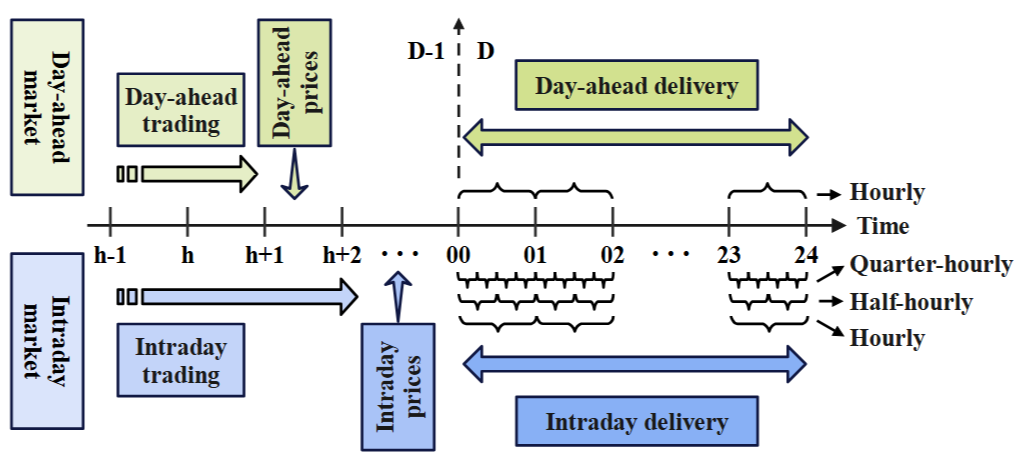}
    \caption{The sequence of events in the DA and ID markets.}
    \label{fig:Markets}
\end{figure}

Chargco participates in sequential DA and ID markets, as depicted in Fig. \ref{fig:Markets}. DA market operates hourly, while ID market operates quarter-hourly\footnote{European ID auctions' market time unit (MTU) varies: hourly in Switzerland, half-hourly in France, and quarter-hourly in Germany, Austria, Belgium, and the Netherlands \cite{epexspot}.}. Chargco optimizes 24 bid curves for the DA market (one per hour) and 96 bid curves for the ID market (one per quarter) to maximize profit.

\vspace{-1mm}

\begin{figure}[!hbt]
    \centering    \includegraphics[width=0.85\columnwidth]{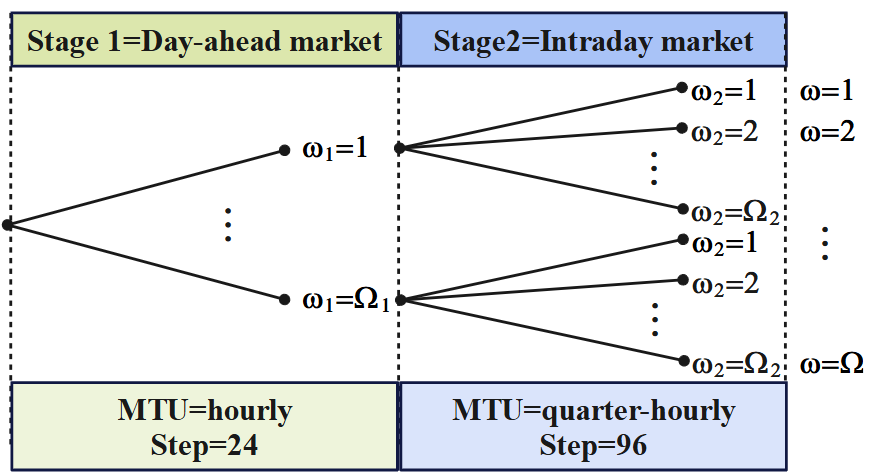}
    \caption{Two-stage stochastic optimization problem.}
    \label{fig:tree}
\end{figure}

The Chargco's profit-maximizing problem is structured as a two-stage stochastic optimization problem, depicted in Fig. \ref{fig:tree}. The first stage involves formulating DA bidding decisions, while the second stage constructs ID bidding curves and schedules equipment operations. This approach explicitly models the impact of ID bids on the DA bids and vice versa.

\vspace{-1mm}

\subsection{Mixed-Integer Quadratic Program (MIQP) formulation}
In this section, we formulate the MIQP model for Chargco's profit-maximization problem over a single day. The objective function, defined in Eq. \eqref{eq.1}, seeks to maximize Chargco's total profit. This profit is defined by calculating the costs of purchasing electricity from both the DA and ID markets and subtracting it from the revenue earned by selling electricity and hydrogen to vehicles. In this formulation, \( t \) represents the time index for the DA market with hourly intervals, and \( t' \) represents the time index for the ID market with quarter-hourly intervals.
\begin{flalign}\label{eq.1}
    & \underset{\Xi}{\mathbf{maximize}} \Big[ \sum_{t} \big(l_{t}^{e} p_{t}^{e} + l_{t}^{h} p_{t}^{h} \big) - \notag\\
    &  \sum_{\substack{t',t\in t', \omega,  \left(\omega_{1},\omega_{2}\right) \in \omega }} \pi_{\omega} \big ( m_{t\omega_{1}}^{d} \xi_{t'\omega_{1}}^{d} + m_{t\omega}^{i} \xi_{t \omega_{2}}^{i} \big) \Big]\\
    &  \Xi = \big[l_{t}^{e},  p_{t}^{e}, l_{t}^{h}, p_{t}^{h}, m_{t\omega_{1}}^{d}, m_{t\omega}^{i}, v_{t\omega}^{e}, v_{t\omega}^{h}, b^{c}_{t\omega}, b^{d}_{t\omega}, b^{l}_{t\omega}, e_{t\omega}^{p},  \notag\\
    & h^{c}_{t\omega}, h^{d}_{t\omega}, h^{l}_{t\omega}, x_{j t'}^{d}, x_{j t \omega_{1}}^{i}, u^{b}_{t\omega}, u^{h}_{t\omega}  \big] \notag &&
\end{flalign}

Equation \eqref{eq.2} ensures that the combined power purchased from both the DA and ID markets matches the electricity allocated for electric vehicle charging, battery charging, and electrolyzer consumption.
\begin{flalign}\label{eq.2}
    & m_{t\omega_{1}}^{d} +m_{t\omega}^{i} = v_{t\omega}^{e} + b^{c}_{t\omega} + e_{t\omega}^{p} \quad  \forall t, \forall \omega, \forall \omega_{1} \in \omega&&
\end{flalign}

Equation \eqref{eq.3} represents the power balance of the battery. Constraints \eqref{eq.4}-\eqref{eq.6} set limits on the battery's inventory level, charging capacity, and discharging capacity, respectively.
\begin{subequations}
\begin{flalign}\label{eq.3}
    & b^{l}_{t\omega}=b^{l}_{t-1,\omega}+ \eta^{b} b^{c}_{t\omega} - b^{d}_{t\omega} / \eta^{b} \; :\tau_{t\omega} \quad \forall t, \forall \omega \\
    \label{eq.4}
    & \underline{b}^{l} \leq b^{l}_{t\omega} \leq \overline{b}^{l}\quad \forall t, \forall \omega \\
    \label{eq.5}
    & \underline{b}^{c} u^{b}_{t\omega} \leq b^{c}_{t\omega} \leq \overline{b}^{c} u^{b}_{t\omega} \; :\underline{\tau}_{t\omega}^{c}, \overline{\tau}_{t\omega}^{c} \quad \forall t, \forall \omega \\
    \label{eq.6}
    & \underline{b}^{d} \left( 1- u^{b}_{t\omega} \right) \leq b^{d}_{t\omega} \leq \overline{b}^{d} \left( 1- u^{b}_{t\omega} \right) \; :\underline{\tau}_{t\omega}^{d}, \overline{\tau}_{t\omega}^{d} \notag\\
    & \forall t, \forall \omega &&
\end{flalign}
\end{subequations}

Equation \eqref{eq.7h} aligns produced hydrogen with tank charging and vehicle refueling. Constraint \eqref{eq.8h} restricts electrolyzer power consumption.
\begin{subequations}
\begin{flalign}\label{eq.7h}
    & \eta^{e} e_{t\omega}^{p}/ \mathrm{H} = h_{t\omega}^{c}+v_{t\omega}^{h}  \quad \forall t, \forall \omega \\
    \label{eq.8h}
    & \underline{e}^{p} \leq e_{t\omega}^{p} \leq \overline{e}^{p} \quad \forall t, \forall \omega &&
\end{flalign}
\end{subequations}

Equation \eqref{eq.3h} outlines hydrogen balance in the tank. Constraints \eqref{eq.4h} to \eqref{eq.6h} limit inventory, charging, and discharging capacity.
\begin{subequations}
\begin{flalign}\label{eq.3h}
    & h^{l}_{t\omega}=h^{l}_{t-1,\omega}+ \eta^{h} h^{c}_{t\omega} - h^{d}_{t\omega} / \eta^{h}  \; :\nu_{t\omega} \quad \forall t, \forall \omega \\
    \label{eq.4h}
    & \underline{h}^{l} \leq h^{l}_{t\omega} \leq \overline{h}^{l}\quad \forall t, \forall \omega \\
    \label{eq.5h}
    & \underline{h}^{c} u^{h}_{t\omega} \leq h^{c}_{t\omega} \leq \overline{h}^{c} u^{h}_{t\omega} \; :\underline{\nu}_{t\omega}^{c}, \overline{\nu}_{t\omega}^{c} \quad \forall t, \forall \omega \\
    \label{eq.6h}
    & \underline{h}^{d} \left( 1- u^{h}_{t\omega} \right) \leq h^{d}_{t\omega} \leq \overline{h}^{d} \left( 1- u^{h}_{t\omega} \right) \; :\underline{\nu}_{t\omega}^{d}, \overline{\nu}_{t\omega}^{d} \notag\\
    & \forall t, \forall \omega &&
\end{flalign}
\end{subequations}

Equation \eqref{eq.7} states the electric vehicle load equals the sum of power purchased and discharged from the battery, while \eqref{eq.17h} ensures hydrogen supplied from the electrolyzer and released from the tank meets hydrogen demand.
\begin{subequations}
\begin{flalign}\label{eq.7}
    & v_{t\omega}^{e} + b^{d}_{t\omega} = l_{t}^{e} \quad \forall t, \forall \omega \\
    \label{eq.17h}
    & v_{t\omega}^{h}+h_{t\omega}^{d} = l_{t}^{h} \quad \forall t, \forall \omega &&
\end{flalign}
\end{subequations}

Equation \eqref{eq.8} specifies DA market bidding curves using a piece-wise linear model, while constraint \eqref{eq.9} ensures their non-increasing nature according to market rules.
\begin{subequations}
\begin{flalign}\label{eq.8}
    & \sum_{t\in t'} m_{t\omega_{1}}^{d} =\frac{\xi_{t'\omega_{1}}^{d}-y_{j}}{y_{j+1}-y_{j}} x_{j+1,t'}^{d} + \frac{y_{j+1}-\xi_{t'\omega_{1}}^{d}}{y_{j+1}-y_{j}} x_{j t'}^{d} \notag\\
    & \text{if} \; y_{j}\leq \xi_{t'\omega_{1}}^{d} < y_{j+1} \quad \forall t', \forall \omega_{1}, \forall j \in \{1,...,J-1 \}
 \\
    \label{eq.9}
    & x_{j+1,t'}^{d} \leq x_{j t'}^{d} \quad \forall t', \forall j \in \{1,...,J-1 \} &&
\end{flalign}
\end{subequations}

Equation \eqref{eq.22} specifies ID market bidding curves, while constraint \eqref{eq.24} ensures their non-increasing nature, based on observed DA market outcomes.
\begin{subequations}
\begin{flalign}\label{eq.22}
    & m_{t\omega}^{i} =\frac{\xi_{t \omega_{2}}^{i}-y_{j}}{y_{j+1}-y_{j}} x_{j+1,t \omega_{1}}^{i} + \frac{y_{j+1}-\xi_{t \omega_{2}}^{i}}{y_{j+1}-y_{j}} x_{j t \omega_{1}}^{i} \notag\\
    & \text{if} \; y_{j}\leq \xi_{t \omega_{2}}^{i} < y_{j+1} \quad \forall t, \forall \omega,(\omega_{1},\omega_{2}) \in \omega \notag\\
    &  \forall j \in \{1,...,J-1 \} \\
    \label{eq.24}
    & x_{j+1,t \omega_{1}}^{i} \leq x_{j t \omega_{1}}^{i} \quad \forall t, \forall \omega_{1}, \forall j \in \{1,...,J-1 \} &&
\end{flalign}
\end{subequations}

In our study, we employ random forest as a surrogate model to capture the relationship between electricity and hydrogen load levels and their selling prices efficiently. Random forest employs bootstrap aggregating to train decision trees on \( B \) random subsets of the data with replacement, utilizing feature bagging during construction. Predictions for new samples \( x' \) are made by averaging predictions from all individual trees \( f_b \) \cite{wang2022robust}, as represented by Equation \eqref{eq.r3}:
\begin{flalign}\label{eq.r3}
&\hat{f} = \frac{1}{B} \sum_{b=1}^{B} f_b(x') &&
\end{flalign}

\begin{figure}[!hbt]
    \centering    \includegraphics[width=0.6\columnwidth]{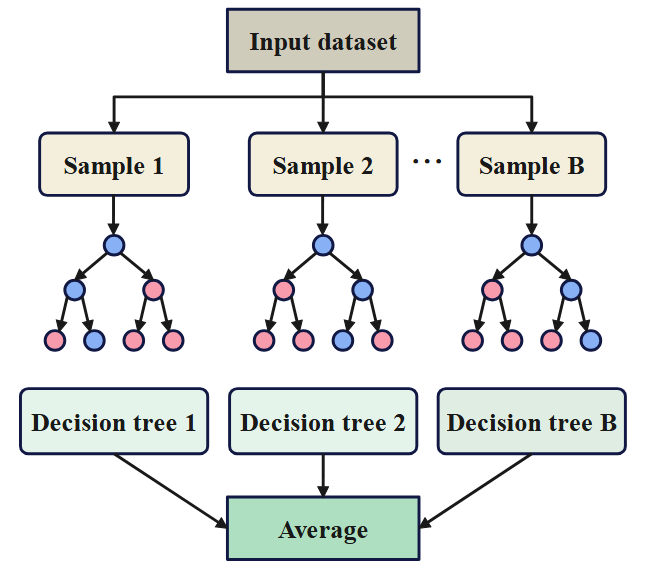}
    \caption{Schematic diagram of the random forest algorithm.}
    \label{fig:random_forest}
\end{figure}

Since increasing the depth of the trees adds complexity to our model, we employed a hybrid approach by combining a shallower random forest regression with a simple linear regression. In this method, each terminal node of the random forest is further refined through a linear regression to enhance predictive accuracy.

To integrate the proposed hybrid machine-learning approach into our MILP model, we encounter multiple linear regression functions corresponding to various segments of the electricity price. Therefore, it is required to identify the active function by determining the corresponding active segment. Binary variable \( u_{tm} \) indicates the active segment, while \( \underline{p}_{tm} \) and \( \overline{p}_{tm} \) bound the price. Equation \eqref{eq.j8} ensures only one segment is active per \( t \), and inequality \eqref{eq.j9} expresses the linear relationship between \( l_{t} \) and \( p_{t} \) within the segment. Parameters are obtained from machine learning, and \( M \) is a large positive constant.
\begin{subequations}
\begin{flalign}\label{eq.j6}
    & \underline{p}_{tm} - \left(1 - u_{tm} \right) M \leq p_{t} \quad \forall t, \forall m \\
    \label{eq.j7}
    & p_{t} \leq \overline{p}_{tm} + \left(1 - u_{tm} \right) M \quad \forall t, \forall m \\
    \label{eq.j8}
    & \sum_{m} u_{tm} = 1 \quad \forall t\\
    \label{eq.j9}
    & l_{t} \leq A_{tm} p_{t} + B_{tm} + \left(1 - u_{tm} \right) M \quad \forall t, \forall m  &&
\end{flalign}
\end{subequations}

The decision variables for the problem are outlined in \eqref{eq.i1} and \eqref{eq.i2}. Lagrange multipliers, corresponding to each constraint, are  separated by colons.
\begin{subequations}
\begin{flalign}\label{eq.i1}
    & u^{b}_{t\omega}, u^{h}_{t\omega} \in \{0, 1 \} \\
    \label{eq.i2}
    & l_{t}^{e},  p_{t}^{e}, l_{t}^{h}, p_{t}^{h}, m_{t\omega_{1}}^{d}, m_{t\omega}^{i}, v_{t\omega}^{e}, v_{t\omega}^{h}, b^{c}_{t\omega}, b^{d}_{t\omega}, b^{l}_{t\omega}, e_{t\omega}^{p}, \notag\\
    & h^{c}_{t\omega}, h^{d}_{t\omega}, h^{l}_{t\omega},  x_{j t'}^{d}, x_{j t \omega_{1}}^{i} \in \mathbb{R}&&
\end{flalign}
\end{subequations}

The derived MIQP is set out below: 
\begin{subequations} \label{MIQP:1}
\begin{flalign}
& \underset{\Xi}{\mathbf{maximize}}~(\ref{eq.1}) \\ 
& \mathbf{Subject~to}: (\ref{eq.2})-(\ref{eq.i2}) &&
\end{flalign}
\end{subequations}

\subsection{Converting MIQP \eqref{eq.1} to a MILP}
For the bilinear terms $l_{t}^{e} p_{t}^{e}$ and $l_{t}^{h} p_{t}^{h}$ in objective function (\ref{eq.1}), we use a piecewise linearization technique with SOS2 variables and a max-affine function. 
Using SOS2 variables is a well-established method for approximating nonlinear functions through a series of line segments. In this approach, only two consecutive variables from the set can take non-zero values, enabling smooth interpolation between predefined breakpoints.
This transforms the original objective function  \eqref{eq.1} to a modified form \eqref{eq.j13}.
\begin{flalign}\label{eq.j13}
    & \underset{\Xi}{\mathbf{maximize}} \Big[ \sum_{t} \big(w_{1t}^{e} - w_{2t}^{e} + w_{1t}^{h} - w_{2t}^{h} \big) - \notag\\
    &  \sum_{\substack{t',t\in t', \omega,  \left(\omega_{1},\omega_{2}\right) \in \omega }} \pi_{\omega} \big ( m_{t\omega_{1}}^{d} \xi_{t'\omega_{1}}^{d} + m_{t\omega}^{i} \xi_{t \omega_{2}}^{i} \big) \Big] &&
\end{flalign}

Since the linearization procedure for both bilinear terms is the same, we express it once as written in \eqref{eq.j10}-\eqref{eq.j4} without superscript $e$ and $h$.
The calculation of \( w_{1t} \) and \( w_{2t} \) is shown in \eqref{eq.j10}. \eqref{eq.j12} demonstrates that the difference between \( w_{1t} \) and \( w_{2t} \) is equal to the product of \( l_{t} \) and \( p_{t} \).
\begin{subequations}
\begin{flalign}\label{eq.j10}
    & w_{1t} = \left(0.5 \left(l_{t} + p_{t} \right)\right)^{2}, w_{2t} = \left(0.5 \left(l_{t} - p_{t} \right)\right)^{2} \quad \forall t \\
    \label{eq.j12}
    & w_{1t} - w_{2t} = l_{t} p_{t} \quad \forall t &&
\end{flalign}
\end{subequations}

\vspace{-1mm}

\subsubsection{Linearization of $\left(0.5 \left(l_{t} + p_{t} \right)\right)^{2}$} 

In equations \eqref{eq.j15}-\eqref{eq.j3}, we compute \( w_{1t} \). Equation \eqref{eq.j15} determines the interval size for \( f_{tm} \) calculation based on \( 0.5 (l_{t} + p_{t}) \) range. Equation \eqref{eq.j14} computes \( m \)-th \( f_{tm} \) value. Equation \eqref{eq.j1} represents \( l_{t} \) and \( p_{t} \) relationship, where \( \beta_{tm} \) is segment contribution. Equation \eqref{eq.j2} calculates \( w_{1t} \), and \eqref{eq.j3} ensures \( \beta_{tm} \) sum equals 1 and are SOS2 type.
\begin{subequations}
\begin{flalign}
    \label{eq.j15}
    & \text{interval}_{tm} = \frac{\text{max}_{tm} - \text{min}_{tm}}{M-1} \\
    \label{eq.j14}
    & f_{tm} = \text{min}_{tm} + m \times \text{interval}_{tm}
    \\ \label{eq.j1}
    & 0.5 \left(l_{t} + p_{t} \right) = \sum_{m} \beta_{tm} f_{tm} \quad \forall t \\
    \label{eq.j2}
    & w_{1t} = \sum_{m} \beta_{tm} \left(f_{tm}\right)^{2} \quad \forall t \\
    \label{eq.j3}
    & \sum_{m} \beta_{tm} = 1, \beta_{tm} \in \text{SOS2} \quad \forall t &&
\end{flalign}
\end{subequations}

\subsubsection{Linearization of $\left(0.5 \left(l_{t} - p_{t} \right)\right)^{2}$} In equation \eqref{eq.j4}, we describe the calculation of \( w_{2t} \) using a max-affine function. Here, we determine the interval size for computing \( f_{tm} \) based on the range between the maximum and minimum values of \( 0.5 (l_{t} - p_{t}) \). Next, we obtain the coefficients \( a_{tm} \) and the constants \( b_{tm} \) for each line. 
\begin{flalign}
    \label{eq.j4}
    & w_{2t} \geq 0.5 \left(l_{t} - p_{t} \right) a_{tm} + b_{tm} \quad \forall t, m=1,...,M-1 &&
\end{flalign}

We prefer the max-affine function to linearize $\left(0.5 \left(l_{t} - p_{t} \right)\right)^{2}$ to avoid introducing SOS2 variables, which would complicate the problem. However, we can't use the max-affine function for $\left(0.5 \left(l_{t} + p_{t} \right)\right)^{2}$ linearization because we maximize a convex problem.
The final stochastic MILP is presented below:
\begin{subequations} \label{MILP:1}
\begin{flalign}
& \underset{\Xi}{\mathbf{maximize}}~(\ref{eq.j13}) \\ 
& \mathbf{Subject~to}: (\ref{eq.2})-(\ref{eq.i2}), (\ref{eq.j6})-(\ref{eq.j4}) &&
\end{flalign}
\end{subequations}

\vspace{-1mm}

\section{Generative Adversarial Network} \label{Generative adversarial network}
The MILP (\ref{MILP:1}) is a stochastic two-stage optimization problem. We use GANs to cluster scenarios representing uncertainties in DA and ID electricity prices, refining this with the Interquartile Range (IQR) to remove outliers. The IQR identifies abnormal outliers beyond 1.5 times its value \cite{yang2023influences}.
\begin{subequations}
\begin{flalign}\label{eq.o1}
& \text{IQR} : \text{Q3} - \text{Q1} \\ \label{eq.o2}
& \text{OU} > \text{Q3} + 1.5 \times \text{IQR}, \; \text{OL} < \text{Q1} - 1.5 \times \text{IQR}
 &&
\end{flalign}
\end{subequations}

\begin{figure}[!hbt]
    \centering    \includegraphics[width=0.9\columnwidth]{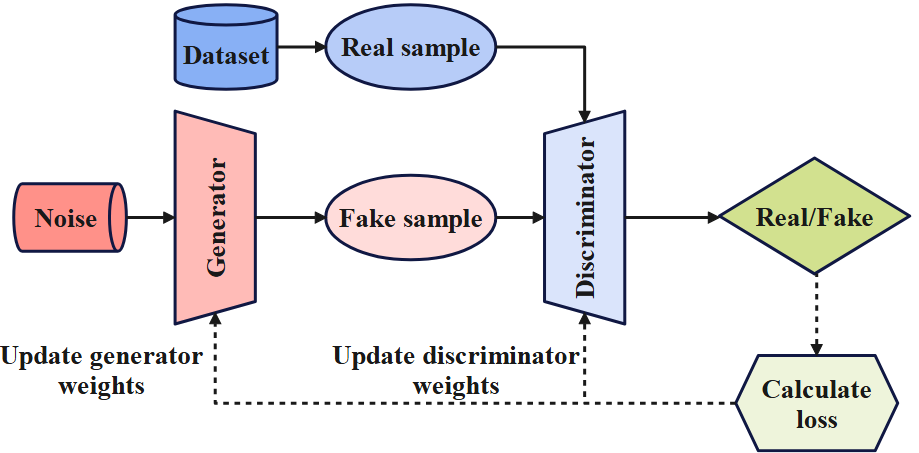}
    \caption{Generative adversarial network structure.}
    \label{fig:GAN}
\end{figure}

The GANs model (Fig. \ref{fig:GAN}) consists of two neural networks: the generator $G(z; l^{G})$ and the discriminator $D(x; l^{D})$, with associated weights $l^{G}$ and $l^{D}$. The generator uses random noise inputs $z$ from distribution $\mathbb{P}_z$ to produce scenarios until the discriminator cannot distinguish them from real historical samples $x$. The discriminator aims to differentiate between real data from $\mathbb{P}_x$ and generated data from $\mathbb{P}_z$. Loss functions $L^{G}$ and $L^{D}$ optimize the neural network weights, creating the min-max optimization model (\ref{eq.g3}). We employed the Keras backend \cite{keras-doc} with TensorFlow \cite{tensorflow} for GANs model development and training.
\begin{subequations}
\begin{flalign}\label{eq.g1}
&\min_{l^{G}} L^{G} = -\mathbb{E}_{z \sim \mathbb{P}_z}\left[\log(D(G(z; l^{G}); l^{D}))\right]\\
\label{eq.g2}
&\min_{l^{D}} L^{D} = -\mathbb{E}_{x \sim \mathbb{P}_x}\left[\log(D(x; l^{D}))\right] - \notag\\
&\mathbb{E}_{z \sim \mathbb{P}_z}\left[\log(1 - D(G(z; l^{G}); l^{D}))\right]\\
\label{eq.g3}
&\min_{l^{G}} \max_{l^{D}} \mathbb{E}_{x \sim \mathbb{P}_x}\left[\log(D(x; l^{D}))\right] + \notag\\
&\mathbb{E}_{z \sim \mathbb{P}_z}\left[\log(1 - D(G(z; l^{G}); l^{D}))\right]&&
\end{flalign}
\end{subequations}

The scenario number is optimal if the Mean Squared Error (MSE) between real and generated data falls below a threshold \(\epsilon\). Equation \eqref{eq.k2} computes the MSE between observed (\(y_i\)) and predicted (\(\hat{y}_i\)) values with \(n\) data points \cite{ashraf2020novel}. As the exact Probability Density Function (PDF) is unknown, we estimate it using Kernel Density Estimation (KDE) due to discrete samples. Equation \eqref{eq.k1} presents the kernel density estimate \(\hat{f}(x)\) for a dataset \(\{X_i\}\), where \(K(\cdot)\) is the kernel function, \(h\) is the bandwidth, and \(n\) is the number of data points \cite{duan2019estimation}.
\begin{subequations}
\begin{flalign}\label{eq.k2}
&\text{MSE} = \frac{1}{n} \sum_{i=1}^{n} (y_i - \hat{y}_i)^2\\
\label{eq.k1}
    & \hat{f}(x) = \frac{1}{n h} \sum_{i=1}^{n} K\left(\frac{x - X_i}{h}\right) &&
\end{flalign}
\end{subequations}

\vspace{-1mm}

\section{The Improved L-Shaped Decomposition (ILSD) Algorithm } \label{Methodology}
Solving the MILP (\ref{MILP:1}) is computationally challenging due to the interconnection of DA and ID markets and a large number of operational variables. To address this, we improve and employ the L-shaped decomposition algorithm, which decomposes the original MILP into MP and SP, iteratively solving them to find the optimal solution. \footnote{In this paper, the optimal solution refers to an \(\epsilon\)-optimal solution where \(\epsilon\) is a small number indicating the optimality gap of the solution. It is defined as \(\epsilon = \frac{UB - LB}{UB}\).}

\vspace{-1mm}

\subsection{The L-shaped decomposition Master Problem}
The electricity and hydrogen load levels, their selling prices and DA power purchases are optimized in the L-shaped decomposition MP and then they are passed to the L-shaped decomposition SP. Optimality cut \eqref{eq.2m} iteratively refines MP decisions with \(\alpha\) underestimating the objective function of SP.
\begin{subequations} 
\begin{flalign}\label{eq.1m}
    & \textbf{L-shaped decomposition MP:} \notag\\  
    & \underset{\Xi}{\mathbf{maximize}} \sum_{t} \big(w_{1t}^{e} - w_{2t}^{e} + w_{1t}^{h} - w_{2t}^{h} \big) - \\ \notag 
    &  \sum_{\substack{t',t\in t', \omega,  \omega_{1} \in \omega }} \pi_{\omega} \big ( m_{t\omega_{1}}^{d} \xi_{t'\omega_{1}}^{d} \big) - \alpha \\
    \label{eq.n1}
    &  \Xi = \big[w_{1t}^{e}, w_{2t}^{e}, w_{1t}^{h}, w_{2t}^{h},\beta_{tm}^{e}, \beta_{tm}^{h}, u^{e}_{t m}, u^{h}_{t m}, l_{t}^{e},  p_{t}^{e}, l_{t}^{h}, p_{t}^{h},  \notag\\
    & m_{t\omega_{1}}^{d}, x_{j t'}^{d}, \alpha  \big] \notag\\
    &\mathbf{subject~to: } \eqref{eq.8}, \eqref{eq.9}, (\ref{eq.j6})-(\ref{eq.j4}) \\
    \label{eq.2m}
    & \alpha \geq - \alpha\left( \hat{m}_{nt\omega_{1}}^{d}, \hat{l}_{nt}^{e}, \hat{l}_{nt}^{h} \right) + 
    \sum_{\substack{t, \omega,  \omega_{1} \in \omega }}  \Big[ \left(\hat{m}_{nt\omega_{1}}^{d} -m_{t\omega_{1}}^{d} \right)  \notag \\
    & \lambda_{nt\omega} + 
    \left(\hat{l}_{nt}^{e} -l_{t}^{e} \right) \vartheta_{nt\omega} + \left(\hat{l}_{nt}^{h} -l_{t}^{h} \right) \varrho_{nt\omega} \Big] \quad \forall n,  \\
    \label{eq.n2}
    & u^{e}_{t m}, u^{h}_{t m} \in \{0, 1 \}, \beta_{tm}^{e}, \beta_{tm}^{h} \in \text{SOS2} \quad \forall t\\
    \label{eq.n7}
    & w_{1t}^{e}, w_{2t}^{e}, w_{1t}^{h}, w_{2t}^{h}, l_{t}^{e},  p_{t}^{e}, l_{t}^{h}, p_{t}^{h},  m_{t\omega_{1}}^{d}, x_{j t'}^{d}, \alpha \in \mathbb{R}&&
\end{flalign}
\end{subequations}

\vspace{-1mm}

\subsection{The L-shaped decomposition Sub-problem}
The L-shaped decomposition SP optimizes ID bidding curves and charging-station operation.
\begin{subequations}
\begin{flalign}\label{eq.3m}
    & \textbf{L-shaped decomposition SP:} \notag\\
    & \underset{\Xi}{\mathbf{maximize}} \sum_{t,\omega,\omega_{2} \in \omega} \pi_{\omega} \left( - m_{t\omega}^{i} \xi_{t \omega_{2}}^{i} \right) \\
    &  \Xi = \big[m_{t\omega}^{i}, v_{t\omega}^{e}, v_{t\omega}^{h}, b^{c}_{t\omega}, b^{d}_{t\omega}, b^{l}_{t\omega}, e_{t\omega}^{p},   h^{c}_{t\omega}, h^{d}_{t\omega}, h^{l}_{t\omega}, \notag\\
    & x_{j t \omega_{1}}^{i}, u^{b}_{t\omega}, u^{h}_{t\omega}  \big] \notag\\
    &\mathbf{subject~to: } \eqref{eq.3}-\eqref{eq.6h}, \eqref{eq.22}, \eqref{eq.24}, \eqref{eq.i1} \\
    \label{eq.4m}
    & \hat{m}_{nt\omega_{1}}^{d} = v_{t\omega}^{e} + b^{c}_{t\omega} + e_{t\omega}^{p} - m_{t\omega}^{i}   \; :\lambda_{nt\omega} \quad \forall n, \forall t, \\
    \notag
    & \forall \omega, \forall \omega_{1} \in \omega\\
    \label{eq.5m}
    & v_{t\omega}^{e} + b^{d}_{t\omega} = \hat{l}_{nt}^{e} \; :\vartheta_{nt\omega} \quad \forall n, \forall t, \forall \omega \\
    \label{eq.6m}
    & v_{t\omega}^{h}+h_{t\omega}^{d} = \hat{l}_{nt}^{h} \; :\varrho_{nt\omega} \quad \forall n, \forall t, \forall \omega \\
    \label{eq.n3}
    & m_{t\omega}^{i}, v_{t\omega}^{e}, v_{t\omega}^{h}, b^{c}_{t\omega}, b^{d}_{t\omega}, b^{l}_{t\omega}, e_{t\omega}^{p},  h^{c}_{t\omega}, h^{d}_{t\omega}, h^{l}_{t\omega}, \notag\\
    & x_{j t \omega_{1}}^{i} \in \mathbb{R}&&
\end{flalign}
\end{subequations}

\vspace{-1mm}

The binary variable $u^{b}_{t\omega}$ and $u^{h}_{t\omega}$, categorize the L-shaped decomposition SP as a MILP. Lemma \ref{lem:sp} reformulates the MILP model of L-shaped decomposition SP as an equivalent LP. 

\begin{lemma} \label{lem:sp}
If the binary variables in the L-shaped decomposition SP (\ref{eq.3m})-(\ref{eq.n3}) are removed, the resulting relaxed LP and the original MILP model of L-shaped decomposition SP have the same optimal solution. 
\end{lemma}

\begin{proof}
Suppose we eliminate binary variables, formulating the problem as an LP. Simultaneous charging and discharging occur in the battery and tank ($b^{c}_{t\omega}, b^{d}_{t\omega} > 0$ and $h^{c}_{t\omega}, h^{d}_{t\omega} > 0$). Applying the KKT optimality conditions, we modify \eqref{eq.2} to \eqref{eq.l2}, with stationary conditions given in \eqref{eq.s1} - \eqref{eq.s4}.
\begin{subequations}
\begin{flalign}
    \label{eq.l2}
    & m_{t\omega_{1}}^{d} +m_{t\omega}^{i} + b^{d}_{t\omega} - l_{t}^{e} - b^{c}_{t\omega} -  \mathrm{H} / \eta^{e} (h_{t\omega}^{c} -  h_{t\omega}^{d} + \notag\\
    & l_{t}^{h})   = 0 \; :\gamma_{t\omega} \quad \forall t, \forall \omega, \forall  \omega_{1} \in \omega\\
    \label{eq.s1}
    & -\gamma_{t\omega} + \eta^{b} \tau_{t\omega} + \underline{\tau}_{t\omega}^{c} - \overline{\tau}_{t\omega}^{c} = 0  \quad \forall t, \forall \omega  \\
    \label{eq.s2}
    &  \gamma_{t\omega} - \tau_{t\omega} / \eta^{b} + \underline{\tau}_{t\omega}^{d} - \overline{\tau}_{t\omega}^{d} = 0 \quad \forall t, \forall \omega && \\
    \label{eq.s3}
    & - \gamma_{t\omega} \mathrm{H}/ \eta^{e} + \eta^{h} \nu_{t\omega} + \underline{\nu}_{t\omega}^{c} - \overline{\nu}_{t\omega}^{c}= 0 \quad \forall t, \forall \omega  \\
    \label{eq.s4}
    & \gamma_{t\omega} \mathrm{H}/ \eta^{e} - \nu_{t\omega} / \eta^{h} + \underline{\nu}_{t\omega}^{d} - \overline{\nu}_{t\omega}^{d}= 0 \quad \forall t, \forall \omega  &&
    \end{flalign}
\end{subequations}

Given the minimum charge and discharge capacities of zero, the Lagrangian multipliers $\underline{\tau}_{t\omega}^{c}$, $\underline{\tau}_{t\omega}^{d}$, $\underline{\nu}_{t\omega}^{c}$, and $\underline{\nu}_{t\omega}^{d}$ are all determined to be zero. Consequently, we obtain \eqref{eq.s5} and \eqref{eq.s6}, leading to the derivation of \eqref{eq.s7}.
\begin{subequations}
    \begin{flalign}
    \label{eq.s5}
    & (\gamma_{t\omega} + \overline{\tau}_{t\omega}^{c} ) / \eta^{b} = \eta^{b} (\gamma_{t\omega} - \overline{\tau}_{t\omega}^{d}) \quad \forall t, \forall \omega  &&
    \\
    \label{eq.s6}
    & ( \gamma_{t\omega} \mathrm{H}/ \eta^{e} + \overline{\nu}_{t\omega}^{c} ) / \eta^{h} = \eta^{h} (\gamma_{t\omega} \mathrm{H}/ \eta^{e} - \overline{\nu}_{t\omega}^{d}) \notag\\
    & \forall t, \forall \omega  \\
    \label{eq.s7}
    & \gamma_{t\omega} 
    (\eta^{b} - 1/ \eta^{b}) + \gamma_{t\omega} \mathrm{H}/ \eta^{e} 
    (\eta^{h} - 1/ \eta^{h}) = \notag\\
    & \overline{\tau}_{t\omega}^{c} / \eta^{b} + \eta^{b} \overline{\tau}_{t\omega}^{d} + \overline{\nu}_{t\omega}^{c} / \eta^{h} + \eta^{h} \overline{\nu}_{t\omega}^{d}\quad \forall t, \forall \omega  &&
\end{flalign}
\end{subequations}

Assuming $ b^{c}_{t\omega}, b^{d}_{t\omega} > 0$ and $ h^{c}_{t\omega}, h^{d}_{t\omega} > 0$, the right-hand side of \eqref{eq.s7} will be either 0 or strictly positive, while the left-hand side expression is negative. This contradiction implies that at least one of the variables $ b^{c}_{t\omega}$ or $ b^{d}_{t\omega}$ and $ h^{c}_{t\omega}$ or $ h^{d}_{t\omega}$ must be zero.
\end{proof}

The L-shaped decomposition algorithm terminates upon meeting convergence criteria based on the relative gap or maximum iterations. The objective function of MP serves as the upper bound, while the total revenue minus DA expenses and SP objective value acts as the lower bound. 

\vspace{-1mm}

\subsection{Improved L-shaped decomposition (ILSD) algorithm}
\begin{figure}[hbt!]
    \centering
\includegraphics[width=0.8\columnwidth]{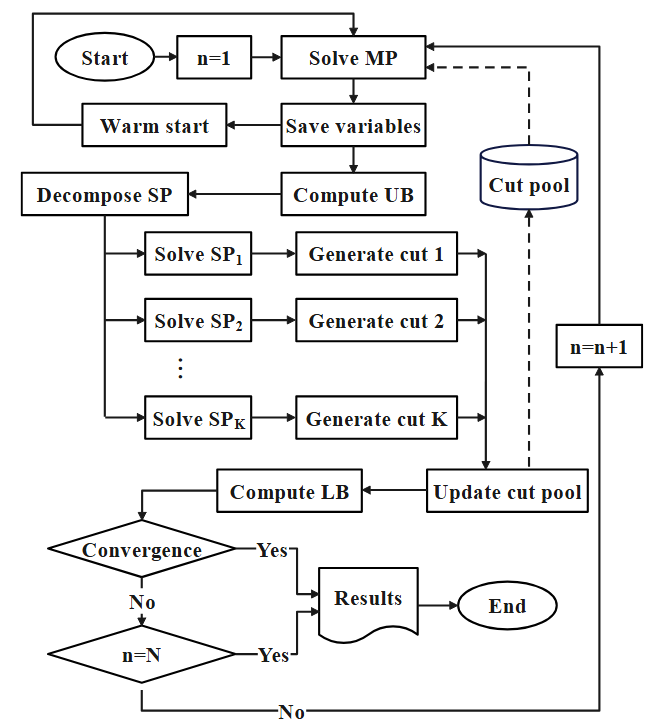}
    \caption{Flowchart of the ILSD algorithm.}
    \label{fig:Algorithm}
\end{figure}

To improve our L-shaped decomposition algorithm, we utilize warm start technique and innovative approach to tackle SP infeasibility. We also customize valid inequalities and employ cut disaggregation methods for enhanced efficiency. Details on implementing these strategies are provided in the subsequent section, along with a flowchart of our proposed algorithm (Fig. \ref{fig:Algorithm}).

\vspace{-1mm}

\subsubsection{Warm Start}
To accelerate our L-shaped implementation, we customized a warm start approach. In each iteration,  MP uses information from the previous iteration's solution.

\subsubsection{Resolve Infeasibility}
When over-procuring power from the DA market renders SP infeasible, solutions include adding slack variables or feasibility cuts. However, slack variables lead to numerical instability, while feasibility cuts requires solving SP again to obtain the extreme ray, which is time-consuming. Therefore, we propose efficiently modifying \eqref{eq.7} to \eqref{eq.7f} to maintain feasibility, effectively penalizing power over-procurement for direct electric vehicle charging.
\begin{flalign}\label{eq.7f}
    v_{t\omega}^{e} + b^{d}_{t\omega} \geq l_{t}^{e} \quad \forall t, \forall \omega &&
\end{flalign}

\begin{lemma}\label{lem:infs}
The relaxed constraint \eqref{eq.7f} is always tight at the optimal solution of optimization problem \eqref{eq.1} - \eqref{eq.i2}. 
\end{lemma}

\begin{proof}
Let \(x^\ast\) be the optimal solution of \eqref{eq.1} - \eqref{eq.i2}, and suppose there exists a set ($\hat{t},\hat{\omega}$) such that \(v_{\hat{t}\hat{\omega}}^{e^*} + b^{d^*}_{\hat{t}\hat{\omega}} > l_{\hat{t}}^{e}\):
\begin{flalign}\label{eq.p2}
    \sum_{t} \left ( v_{t\hat{\omega}}^{e^*} + b^{d^*}_{t\hat{\omega}} \right) > \sum_{t} l_{t}^{e} \quad \exists \; \hat{\omega}&&
\end{flalign}

By incorporating \eqref{eq.2}, \eqref{eq.p2} transforms into \eqref{eq.p4}, where $\phi_{\hat{\omega}}^*$ is a positive auxiliary variable for \eqref{eq.p2}.
\begin{flalign}
    \label{eq.p4}
    & \sum_{t} \left (m_{t\hat{\omega}_{1}}^{d^*} +m_{t\hat{\omega}}^{i^*} \right) = \sum_{t} \left ( b^{c^*}_{t\hat{\omega}} + e_{t\hat{\omega}}^{p^*} - b^{d^*}_{t\hat{\omega}}  + l_{t}^{e} \right) + \notag\\
    & \phi_{\hat{\omega}}^* \quad \exists \; \hat{\omega}, \hat{\omega}_{1} \in \hat{\omega}&&
\end{flalign}

The optimal objective function $\mathcal{F} (x^*)$ is given by \eqref{eq.p5}, where $\phi_{\hat{\omega}}^*$ and $ \sum_{t} \left ( b^{c^*}_{t\hat{\omega}} + e_{t\hat{\omega}}^{p^*} - b^{d^*}_{t\hat{\omega}}  + l_{t}^{e} \right)$ are equal to $ \phi_{\hat{\omega}}^{1^*} + \phi_{\hat{\omega}}^{2^*}$ and   
$\sum_{t} \left (\mu_{t \hat{\omega}}^{1^*} + \mu_{t\hat{\omega}}^{2^*} \right)$, respectively.
\begin{flalign}\label{eq.p5}
    &  \mathcal{F} (x^*)=\sum_{t} \big(l_{t}^{e} p_{t}^{e} + l_{t}^{h} p_{t}^{h} \big) -\sum_{\substack{t',t \in t', \omega,   \\ \left(\omega_{1},\omega_{2}\right) \in \omega, \omega \neq \hat{\omega} }} \pi_{\omega} \big ( m_{t\omega_{1}}^{d} \xi_{t'\omega_{1}}^{d} + \notag\\
    & m_{t\omega}^{i} \xi_{t \omega_{2}}^{i} \big) -  \pi_{\hat{\omega}} \big( \phi_{\hat{\omega}}^{1^*} \xi_{\hat{t}\hat{\omega}_{1}}^{d}+ \sum_{\substack{t',t \in t', \hat{\omega_{1}} \in \hat{\omega}}} \mu^{1^*}_{t\hat{\omega}} \xi_{t'\hat{\omega}_{1}}^{d}  \big) - \notag\\
    & \pi_{\hat{\omega}} \big(\phi_{\hat{\omega}}^{2^*} \xi_{\hat{t}\hat{\omega}_{2}}^{i}+\sum_{\substack{t',t \in t', \hat{\omega_{2}} \in \hat{\omega}}} \mu^{2^*}_{t\hat{\omega}} \xi_{t\hat{\omega}_{2}}^{i} \big) && 
\end{flalign}

By tending $\phi_{\hat{\omega}}^*$ to zero, a new feasible solution $x'$ is obtained where \(v_{\hat{t}\hat{\omega}}^{e} + b^{d}_{\hat{t}\hat{\omega}} = l_{\hat{t}}^{e}\) and $\mathcal{F} (x')$ is as follows:
\begin{flalign}\label{eq.p8}
    & \mathcal{F} (x') = \mathcal{F} (x^*) + \pi_{\hat{\omega}} \xi_{\hat{t}\hat{\omega}_{1}}^{d} \phi_{\hat{\omega}}^{1^*} + \pi_{\hat{\omega}} \xi_{\hat{t}\hat{\omega}_{1}}^{i} \phi_{\hat{\omega}}^{2^*}&&
\end{flalign}

Since $\pi_{\hat{\omega}} \xi_{\hat{t}\hat{\omega}_{1}}^{d} \phi_{\hat{\omega}}^{1^*} + \pi_{\hat{\omega}} \xi_{\hat{t}\hat{\omega}_{1}}^{i} \phi_{\hat{\omega}}^{2^*} > 0$, then $\mathcal{F} (x') > \mathcal{F} (x^*)$, which is in contrast with the optimality of $x^*$. 
\end{proof}

Using constraint \eqref{eq.7f}, we transform equation \eqref{eq.4m} into constraint \eqref{eq.4fm}, expediting the L-shaped decomposition process.
\begin{flalign}\label{eq.4fm}
    & \hat{m}_{nt\omega_{1}}^{d} - \hat{l}_{nt}^{e} \geq - b^{d}_{t\omega} + b^{c}_{t\omega} + e_{t\omega}^{p} - m_{t\omega}^{i}   \; :\lambda_{nt\omega} \\ \notag
    & \forall n, \forall t, \forall \omega,
     \forall \omega_{1} \in \omega&&
\end{flalign}

\subsubsection{Valid inequalities}
We introduce valid inequality \eqref{eq.2v} into the MP, which quickly removes infeasible solutions and accelerates convergence. This is necessary because our MP lacks constraints to prevent excessive power procurement from exceeding total demand, battery, and electrolyzer capacity.
\begin{flalign}\label{eq.2v}
    m_{t\omega_{1}}^{d} \leq l_{t}^{e} + \overline{b}^{c} + \overline{e}^{p}  \quad \forall t, \forall \omega_{1}&&
\end{flalign}

\subsubsection{Disaggregation of optimality cut}
We adopt a practical approach to balance iteration quantity and cut production by generating several cuts during each iteration, transitioning from \eqref{eq.2m} to \eqref{eq.1mn}, where $k$ denotes the number of cuts.
\begin{subequations} 
\begin{flalign}\label{eq.1mn}
    & \alpha_{k}\geq - \alpha_{k}\left( \hat{m}_{nt\omega_{1}}^{d}, \hat{l}_{nt}^{e}, \hat{l}_{nt}^{h} \right) + 
    \sum_{\substack{t,\omega, \omega_{1} \in \omega }} \Big[ \left(\hat{m}_{nt\omega_{1}}^{d} -m_{t\omega_{1}}^{d} \right)   \notag \\
    & 
    \lambda_{nt\omega} + \left(\hat{l}_{nt}^{e} -l_{t}^{e} \right) \vartheta_{nt\omega} + \left(\hat{l}_{nt}^{h} -l_{t}^{h} \right) \varrho_{nt\omega} \Big] \quad \forall n, \forall k&&
\end{flalign}
\end{subequations}

\begin{figure}[hbt!]
    \centering
\includegraphics[width=0.95\columnwidth]{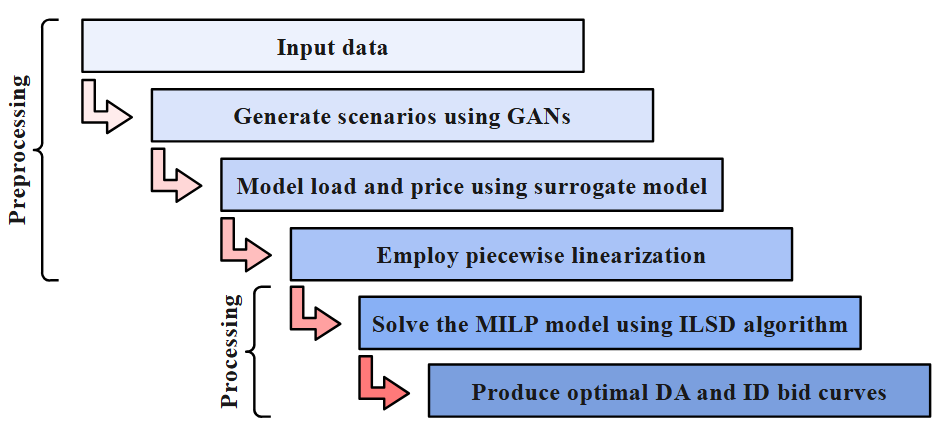}
    \caption{The flowchart of our sequential preprocessing and processing steps.}
    \label{fig:flowchart}
\end{figure}

The flowchart illustrating all the aforementioned components involved in solving our problem is depicted in Fig. \ref{fig:flowchart}. Our approach combines several methods to address specific challenges in optimizing trading for Chargco in day-ahead and intraday markets. We have organized our approach into preprocessing and processing stages to provide a clearer overview. Each of these methods is specifically chosen to address unique challenges regarding market dynamics and its operational characteristics.
\begin{itemize}
    \item \textbf{Preprocessing:}
    \begin{itemize}
        \item \textit{Generative Adversarial Networks (GANs)}: We incorporate GANs to cluster scenarios, effectively managing the uncertainty in electricity prices without increasing computational complexity.
        \item \textit{Surrogate model}: We utilize random forest and linear regression to predict electricity and hydrogen load levels and their selling prices with accuracy and then incorporate them in our MILP formulation. This step is crucial for ensuring robust trading decisions based on reliable data forecasts.
        \item \textit{Linearization of Bilinear Terms}: To handle bilinear terms, we employ piecewise linearization with SOS2 variables and max-affine function. This allows us to approximate nonlinear relationships within the model, maintaining tractability.
    \end{itemize}
    \item \textbf{Processing:}
    \begin{itemize}
        \item \textit{Improved L-Shaped Decomposition (ILSD) Algorithm}: Our ILSD algorithm addresses computational challenges by dividing the problem into a master problem and sub-problem, sequentially optimizing trading strategies. It enhances ILSD computational efficiency using warm-start technique, valid inequalities and multiple cuts.
        \item \textit{Optimal Bidding Curves}: The ILSD algorithm ultimately aids in obtaining optimal bidding curves for day-ahead and intraday market participation.
    \end{itemize}
\end{itemize}

\vspace{-1mm}

\section{Numerical Experiments} \label{Case study}
We consider a Chargco participating in the DA and ID markets. The relevant information is provided in Table \ref{tab:data}. 

\begin{table}[hbt!]
    \centering
    \caption{Required information of devices}
    \label{tab:data}
    \begin{tabular}{l p{2cm} l l}
    \toprule
    Parameter & Value & Parameter & Value \\ \midrule
    $\eta^{b}$ & $0.85$ & $\eta^{h}$ & $0.9$ \\
    $\underline{b}^{l}, \overline{b}^{l}$ & $0, 60 \; \text{(kWh)}$  & $\underline{h}^{l}, \overline{h}^{l}$ & $0, 20 \; \text{(kg)}$ \\
    $\underline{b}^{c}, \overline{b}^{c}$& $0, 15 \; \text{(kW)}$ & $\underline{h}^{c}, \overline{h}^{c}$ & $0, 5 \; \text{(kg/h)}$ \\
    $\underline{b}^{d}, \overline{b}^{d}$& $0, 15 \; \text{(kW)}$ & $\underline{h}^{d}, \overline{h}^{d}$ & $0, 5 \; \text{(kg/h)}$ \\
    $\underline{e}^{p}, \overline{e}^{p}$& $0, 1000 \; \text{(kW)}$ & $\eta^{e}$& $0.8$ \\ \bottomrule
    \end{tabular}
\end{table}

\subsection{Scenario generation}
\begin{figure}[hbt!]
    \centering
\includegraphics[width=0.8\columnwidth]{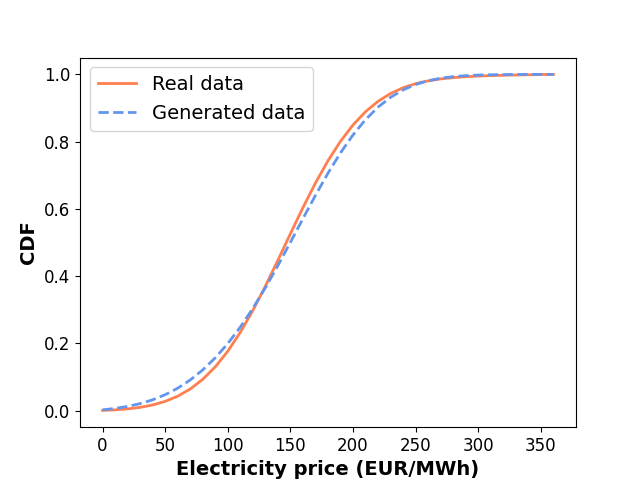}
    \caption{CDF of the original dataset versus generated scenarios by GANs.}
    \label{fig:CDF}
\end{figure}

\begin{figure}[hbt!]
    \centering
\includegraphics[width=0.8\columnwidth]{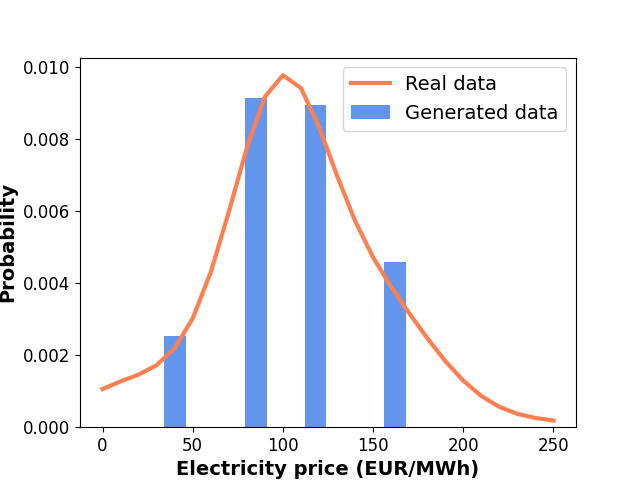}
    \caption{PDF of the original dataset versus generated scenarios by GANs.}
    \label{fig:PDF}
\end{figure}

We evaluate our GANs model's ability to cluster electricity market prices by comparing the Cumulative Distribution Function (CDF) in Fig.~\ref{fig:CDF}. Using both GANs-generated scenarios and data from the German market provided by ENTSO-E \cite{entsoe}, our analysis confirms a close match between the two CDFs, demonstrating accurate capture of electricity price dynamics. In stochastic programming, discrete distributions simplify extensive datasets for estimating probability distributions. Figure \ref{fig:PDF} illustrates this, discretizing a continuous PDF into four scenarios shown as rectangular bars.


\begin{figure}[hbt!]
    \centering
\includegraphics[width=0.8\columnwidth]{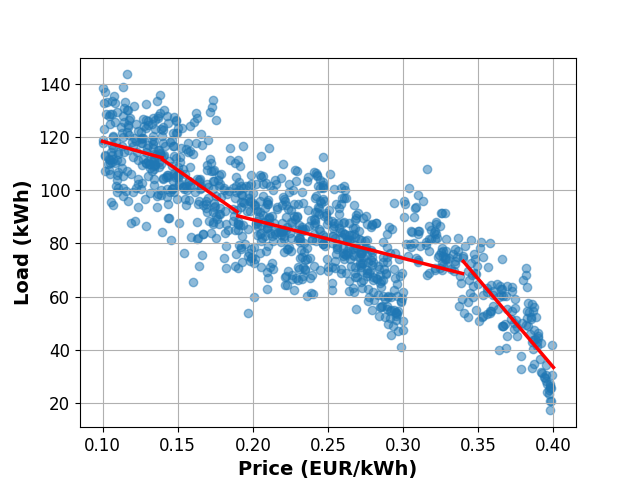}
    \caption{Load and price relationship using random forest and linear regression.}
    \label{fig:load}
\end{figure}

The application of random forest followed by linear regression offers insights into the dependency of electricity and hydrogen load levels on their price variations across different intervals. In Fig. \ref{fig:load}, we illustrate the relationship between electricity load and price for a selected hour as an example.

\vspace{-1mm}

\subsection{Market analysis}
Table \ref{tab:table1} offers operational insights into the charging station dynamics, encompassing key parameters like energy levels, loads, pricing trends, and purchases across different times. It enables analysis of how price changes influence operations. During low market prices, energy is stored, while during high prices, stored energy is discharged to minimize costs and enhance profitability by adapting to market conditions.
\begin{table}[hbt!]
    \centering
    \caption{Charging Station Operation}
    \label{tab:table1}
    \setlength{\tabcolsep}{2.5pt} 
    \begin{tabular}{l *{8}{c}}
    \toprule
        & \textbf{t1} & \textbf{t2} & \textbf{t3} & \textbf{t4} & \textbf{t5} & \textbf{t6} & \textbf{t7} & \textbf{t8} \\
        \midrule
        DA price (EUR/MWh) & 83.6 & 83.6 & 83.6 & 83.6 & 33.3 & 33.3 & 33.3 & 33.3 \\
        ID price (EUR/MWh) & 102.6 & 88.7 & 80.3 & 61.9 & 15.0 & 82.9 & 74.3 & 20.0 \\     
        Battery level (kWh) & - & - & - & - & 12.8 & - & - & - \\
        Hydrogen level (kg) & - & - & - & - & 4.5 & 4.5 & - & - \\
        Electricity load (kWh) & 5.4 & 57.1 & 118.3 & 77.9 & 57.4 & 97.4 & 97.7 & 24.7 \\
        Hydrogen load (kg) & 2.2 & 4.7 & 9.3 & 4.2 & 4.6 & 6.3 & 7.0 & 3.2 \\
        Electricity price (\$/kWh) & 0.4 & 0.4 & 0.4 & 0.4 & 0.4 & 0.4 & 0.4 & 0.3 \\
        Hydrogen price (\$/kg) & 10.8 & 13.0 & 14.0 & 14.0 & 13.0 & 14.0 & 14.0 & 10.6 \\
        DA purchase (kWh) & 111 & 287 & - & - & - & 394 & 244 & - \\
        ID purchase (kWh) & - & - & 576 & 282 & 543 & - & - & 185 \\
        \bottomrule
    \end{tabular}
\end{table}

\begin{figure}[hbt!]
    \centering
\includegraphics[width=0.8\columnwidth]{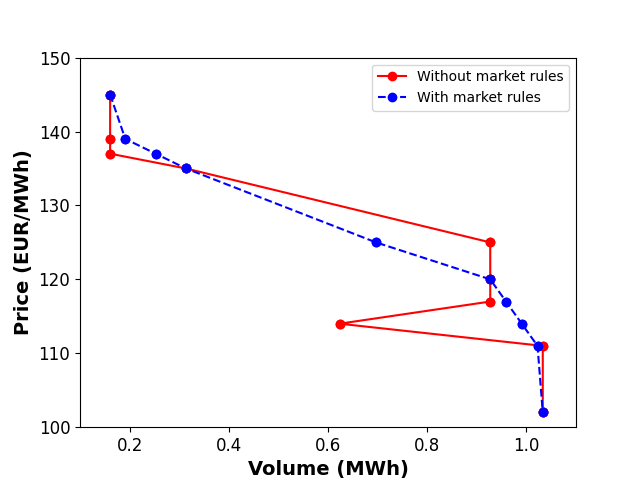}
    \caption{The bidding curves without and with market-rule consideration.}
    \label{fig:bid}
\end{figure}

\begin{figure}[hbt!]
    \centering
\includegraphics[width=0.8\columnwidth]{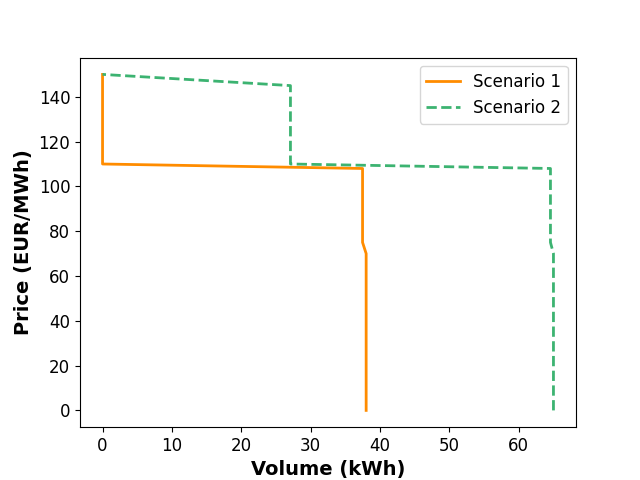}
    \caption{ID bidding curves reflecting two DA market scenarios.}
    \label{fig:bids}
\end{figure}

Figure \ref{fig:bid} illustrates the downward-sloping DA bidding curves (blue curve) for the Chargco for one selected hour. The red curve depicts the bidding curve when the downward-sloping condition is relaxed.
Figure \ref{fig:bids} shows ID bidding curves for one quarter-hour, reflecting two scenarios from the DA market. Different bidding curves in the ID market correspond to realized scenarios from the DA market, highlighting sensitivity to DA market price fluctuations and emphasizing the need for accurate clustering of electricity market prices.

\vspace{-1mm}

\vspace{-1mm}

\subsection{Computational discussions}
\begin{figure}[hbt!]
    \centering
\includegraphics[width=0.9\columnwidth]{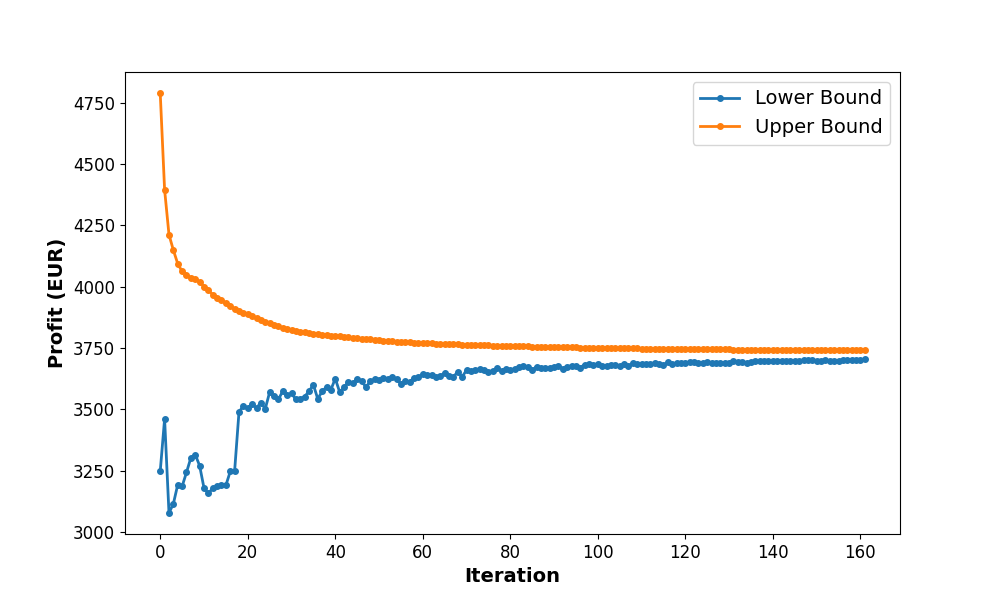}
    \caption{Convergence procedure of our proposed ILSD algorithm.}
    \label{fig:convergence}
\end{figure}

Table \ref{tab:solver} demonstrates that our ILSD algorithm successfully identifies the optimal solution with 2,300 scenarios over 96 discrete 15-minute intervals. In contrast, standard algorithms such as Benders decomposition (BD) and popular solvers like CPLEX and Gurobi fail to provide a solution, even for the linearized formulation, within a time frame of 4 hours. The convergence of our ILSD algorithm, illustrated in Fig. \ref{fig:convergence}, demonstrates the effectiveness of our solution method for optimizing the proposed MILP model of Chargco trading.
\begin{table}[hbt!]
\centering
\caption{THE COMPUTATIONAL RESULTS FOR OUR CASE STUDY}
\label{tab:solver}
\begin{tabular}{ccccc}
\toprule
 & Profit (EUR) & Time (h) & Iteration & Gap (\%) \\
\midrule
CPLEX solver & * & * & * & * \\
Gurobi solver & * & * & * & * \\
BD & * & * & * & * \\
\textbf{Our ILSD} & \textbf{3740} & \textbf{4} & \textbf{162} & \textbf{1} \\
\bottomrule
\end{tabular}

\par\vspace{2mm}
\parbox{0.85\linewidth}{\small Note: * indicates that the solver or method could not solve the problem within the time frame of 4 hours.}
\end{table}

\vspace{-1mm}

\section{Further discussions and future works} \label{Further discussions and future works}

Another complexity in this study arises from the time-decoupled framework. The problem spans 24 hours, and due to intrinsic time coupling in the energy storage constraints, it cannot be solved independently for each time period. This aligns with findings in the literature, such as those by Derakhshandeh et al. \cite{derakhshandeh2017new}, who discuss the challenges associated with time coupling in energy management. The time-coupling introduces an additional layer of complexity to our reformulated MILP model. The time-coupled constraints can be addressed using two main techniques: Dantzig-Wolfe decomposition and Lagrangian relaxation \cite{chakrabarti2023transmission}. In Dantzig-Wolfe decomposition, the time-coupled constraint is eliminated by dividing the problem into a master problem and a subproblem. In Lagrangian relaxation, the time-coupled constraint is dualized in the objective function to create an easier-to-solve problem \cite{martin2012large}.

Recent studies have further explored these techniques. For instance, Shi et al. introduce a two-step decoupling method that effectively separates time-dependent reliability constraints in reliability-based design optimization problems, thereby enhancing computational efficiency in addressing time-dependent uncertainties in engineering design optimization \cite{shi2020novel}.
A multi-stage robust transmission-constrained unit commitment model that employs implicit decision rules and a time-decoupled decomposition framework is presented in \cite{li2019multi}, ensuring robustness, non-anticipative economic dispatch, and scalability for large-scale transmission-constrained unit commitment problems involving uncertain power injections.
A time-decoupled framework for optimizing the multi-period dispatch of distributed energy resources in unbalanced distribution feeders is presented in \cite{nazir2020optimal}. This approach effectively separates temporal dependencies, enhancing computational efficiency while addressing operational constraints across different time periods.
A real-time energy management strategy for flexible traction power supply systems is presented in \cite{zhang2024real}, utilizing a time-decoupled approach to separate time-dependent operational constraints. The proposed approach enhances computational efficiency while effectively scheduling energy resources across different time periods.
Authors in \cite{pourahmadi2020uncertainty} present a framework that evaluates the uncertainty cost of stochastic power producers, focusing on metrics that impact power grid flexibility. This work explores how uncertainty can be effectively managed across different time periods, contributing to a more robust understanding of time-decoupled approaches in energy systems.

As a future work, investigating solution methods for managing time-coupled constraints, such as those associated with energy storage in our MILP problem, would be a valuable direction for research in this study.

\section{Conclusion} \label{Conclusion}
This paper presents a framework for optimizing a Chargco's participation in electricity auction markets. It employs a two-stage stochastic optimization framework, GANs for scenario clustering of electricity prices, and a hybrid approach using random forests and linear regression to depict load-price correlations. We linearize our MIQP using SOS2 variables and a max-affine function. We introduce an ILSD algorithm that effectively tackles L-shaped decomposition SP infeasibilities through a suggested procedure. It incorporates the integration of warm start, valid inequalities and the application of multiple-cuts generation, thereby streamlining computational complexity. We also reformulate the MILP model of L-shaped decomposition SP as an LP, validated by KKT conditions.  
Our proposed optimization algorithm ensures the maximum achievable profit for Chargco by efficiently managing resources, in environments with volatile electricity prices. It enables optimal and timely decision-making, significantly enhancing profitability and operational efficiency as compared to heuristic optimization algorithms.

\vspace{-1mm}

\section*{Acknowledgments}
The research has been supported by the Czech Technical University in Prague, grant number SGS24/094/OHK5/2T/13, and the European Union under the project ROBOPROX (reg. no. CZ.02.01.01/00/22\_008/0004590).

\vspace{-1mm}

\bibliographystyle{IEEEtran}
\bibliography{paper}

\section{Biography Section}

\begin{IEEEbiography}[{\includegraphics[width=1in,height=1.25in,clip,keepaspectratio]{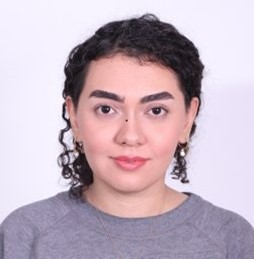}}]{Farnaz Sohrabi}
is a PhD student at the Czech Technical University in Prague. She holds an M.Sc. and a B.Sc. in Electrical Engineering. Her research interests include energy markets, applied mathematics, and operations research.
\end{IEEEbiography}

\begin{IEEEbiography}[{\includegraphics[width=1in,height=1.25in,clip,keepaspectratio]{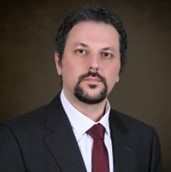}}]{Mohammad Rohaninejad}
is a researcher in the Department of Industrial Informatics at the Institute of Informatics, Robotics, and Cybernetics, Czech Technical University in Prague. He holds a PhD, M.Sc., and B.Sc. in Industrial Engineering, with a focus on applied mathematics and operations research in supply chain, logistics, production planning, and the energy market.
\end{IEEEbiography}

\begin{IEEEbiography}[{\includegraphics[width=1in,height=1.25in,clip,keepaspectratio]{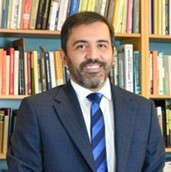}}]{Mohammad Reza Hesamzadeh}
is professor of energy markets at KTH Royal Institute of Technology, Sweden. His areas of interest include energy markets, applied mathematics, and operations research. He is currently in the editorial board of IEEE Transactions on Energy Market, Policy and Regulation.
\end{IEEEbiography}

\begin{IEEEbiography}[{\includegraphics[width=1in,height=1.25in,clip,keepaspectratio]{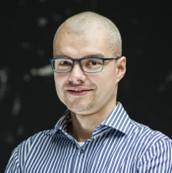}}]{J\'{u}lius~Bem\v{s}}
works as an associate professor at the Department of Economics, Management, and Humanities. He focuses mainly on modern technologies in the power industry, the integration of energy markets, and the economic effectiveness of projects in the power sector.
\end{IEEEbiography}

\end{document}